\documentclass{article}
\usepackage{spconf,amsmath,graphicx,hyperref}
\usepackage{booktabs}
\usepackage{xcolor}

\newcommand{\para}[1]{\par\noindent\textbf{#1}\ }
\title{\vspace{-7.899mm}HEARD BUT NOT SAID: PROMPTING UNDERSTATES TEMPORAL INFORMATION\\OF MUSICAL MOMENTS IN FROZEN LARGE AUDIO-LANGUAGE MODELS}

\name{Eunsik Shin\thanks{Preprint. Under review.}}
\address{Music and Audio Research Group (MARG)\\
Department of Intelligence and Information, Seoul National University\\
esshin@snu.ac.kr}

\begin{document}
\ninept
\maketitle

\begin{abstract}
Large audio-language models (LALMs) understand music, yet struggle to locate a described moment in time. We ask whether this reflects missing temporal information or a bottleneck in how the time span is extracted. Using a paired-query probe built from Slakh2100 stem activations, we attach a \emph{state} query (e.g., ``only piano and bass are playing'') and a \emph{change} query (e.g., ``the guitar drops out, leaving piano and bass'') to the same span and vary how much preceding audio the model is shown. Across frozen LALMs, three ways of extracting the span ask different questions: full-track prompting asking whether the model can say when; windowed yes/no scoring asking whether it verifies the moment when handed the segment, and whether it needs what came before; attention asking whether it can find the moment on its own. \emph{State} queries score above baseline without context, while \emph{change} queries improve with preceding context in most short-segment models; 30\,s padded encoders show no comparable recovery. Prompting and attention expose almost none of this. Prompting failures thus understate what LALMs can verify, though not what they can locate, by an amount set by query type and encoder.
\end{abstract}


\vspace{-0.33em}
\section{Introduction}
\label{sec:intro}
\vspace{-0.43em}
Temporal grounding---returning the time span in a long signal that matches a natural-language description---has been studied across several domains. In video, work has progressed from early proposal-based methods~\cite{gao2017tall,hendricks2017didemo} to DETR-style span predictors~\cite{lei2021momentdetr,moon2023qddetr}. The audio counterpart is more recent: language-based audio moment retrieval~\cite{munakata2025amr} and frame-wise language-audio alignment~\cite{wu2025flam} mostly target environmental sound events, leaving music comparatively unexplored. Temporal grounding in music is particularly challenging because a moment often shares instrumentation, tempo, and other properties with the rest of the track, making localization depend on both local evidence and comparison with surrounding or preceding context.

Large audio-language models (LALMs), which couple an audio encoder to an instruction-following language model~\cite{deshmukh2023pengi,gong2024ltu,tang2024salmonn,chu2024qwen2audio,xu2025qwen25omni,ghosh2025af3}, can be natural candidates for music temporal grounding. Recent work has begun to test this in other domains: activation steering shows that general-audio LALMs carry temporal signal that prompting does not surface~\cite{lin2026steering}; a trained head on internal representations localizes efficiently~\cite{an2026encodeonce}; attention analysis finds layer-specific audio attention during timestamp generation~\cite{listenfirst2026} and a small set of audio-specialist heads indicates when a model actually draws on its audio input~\cite{audioheads2026}. Together, these suggest that prompting may understate temporal information available within LALMs, echoing findings in language models that direct probability measurements can reveal capabilities missed by metalinguistic prompting~\cite{hu2023prompting,gekhman2025insideout,orgad2025know}.

On music, however, LALMs have so far been examined mainly through prompting~\cite{kwon2026mustbench}, while existing probes use coarse event choices~\cite{lin2026steering} or single-instrument notes~\cite{mgbench2026}, leaving it unclear whether models fail to identify what is playing or to detect change relative to preceding context, and how these abilities depend on encoder design. We address this with a controlled, training-free probe. From Slakh2100 instrument stem activity we build spans that can each be described as a \emph{state} query judged from the moment alone or a \emph{change} query that requires comparison with what came before. Both share the audio, position, and duration, so the positional baseline is identical by construction and any difference isolates the temporal computation the query demands. We then hold model, audio, and query fixed and vary how the time span is extracted---prompting for timestamps over the full track, scoring a yes/no judgment window by window under controlled preceding contexts, or reading attention over audio tokens---across seven LALMs whose encoders segment time differently.
All models verify local states above chance when given the relevant segment, while change queries benefit from preceding context mainly in models with short-segment encoders; 30\,s padded encoders do not benefit from added context, while their state scores decline. Meanwhile, full-track prompting and attention expose little of this signal.

This study introduces a paired-query probe that controls audio, position, duration, and chance while manipulating preceding context, and evaluates three training-free ways of extracting temporal span information from frozen LALMs over full-length multi-instrument music. The probe set, prompts, and code will be released upon publication.

\begin{figure*}[t]
\centering
\includegraphics[width=0.55\textwidth]{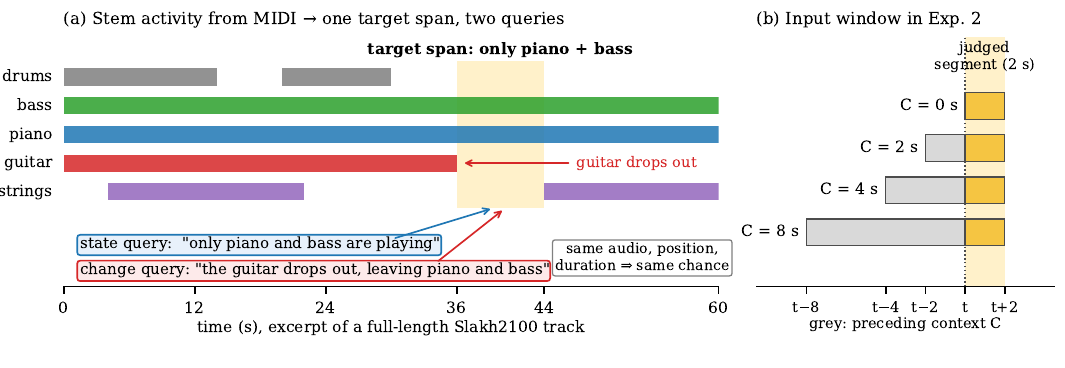}
\caption{(a) Stem activity read from Slakh2100 MIDI. A sparse texture (only piano and bass) that begins right after a stem departure (guitar) becomes one target span with two queries sharing the same audio, so chance is identical. (b) In Exp.~2 the model sees a window that ends with the judged 2\,s segment and begins $C$ seconds earlier; at $C{=}0$ the departure lies outside the window and cannot be verified.}
\label{fig:data}
\end{figure*}
\vspace{-0.63em}

\vspace{-0.80em}
\section{Related Work}
\label{sec:related}
\vspace{-0.63em}
\para{Audio-language models.} Audio-language models pair a pretrained audio encoder with a language model through projection and instruction tuning. Pengi~\cite{deshmukh2023pengi}, LTU~\cite{gong2024ltu}, and SALMONN~\cite{tang2024salmonn} established audio question answering; Qwen2-Audio~\cite{chu2024qwen2audio} and Audio Flamingo 3~\cite{ghosh2025af3} scaled it with Whisper-style encoders~\cite{radford2023whisper}, while Qwen2.5-Omni and Qwen3-Omni~\cite{xu2025qwen25omni,qwen3omni2025} use encoders that process short fixed segments; other models like Kimi-Audio~\cite{kimiaudio2025} and Step-Audio~\cite{stepaudio2025} are general-purpose, while Music Flamingo~\cite{ghosh2026musicflamingo} and MOSS-Music~\cite{mossmusic2026} target music.

\para{Temporal grounding in audio.} AM-DETR defined language-based audio moment retrieval~\cite{munakata2025amr}; FLAM learns frame-wise language-audio alignment~\cite{wu2025flam}; SpotSound adds explicit timestamp grounding to an LALM through training~\cite{sun2026spotsound}. Closest to us, Lin and Lee~\cite{lin2026steering} revealed latent temporal signal that prompting does not surface, but in a three-event setting (chance 27.7\%) without a context manipulation or encoder comparison. Bhattacharya et al.\ benchmark LALM temporal reasoning with confidence estimates~\cite{bhattacharya2025temporal}, and Kulkarni et al.\ trace failures on environmental-sound multiple-choice questions to under-use of audio and mitigate them by attention scaling~\cite{kulkarni2026failure}; both use multiple-choice tasks on environmental sounds (the latter intervening on attention), whereas we score continuous spans in music and compare extraction methods without intervening. Reusing audio-LM internals with a trained head is efficient~\cite{an2026encodeonce} but cannot distinguish information already present from information the head learns; our probe adds no parameters.

\para{Temporal grounding in music.} MusTBench benchmarks temporal reasoning in music LLMs and improves it by training~\cite{kwon2026mustbench}; concurrent work grounds piano MIDI at the note level~\cite{mgbench2026}; music grounding by short video localizes a music moment for a video query~\cite{xin2025mgsv}. We leave models frozen and use text queries over multi-instrument full-length music in which moments share global context.

\vspace{-0.33em}
\section{Paired Probe and Span Extraction}
\label{sec:method}
\vspace{-0.43em}
\subsection{Probe construction}
\label{ssec:data}

Slakh2100~\cite{manilow2019slakh} renders multi-instrument MIDI into audio with one stem per instrument, so we know exactly which instruments sound at every moment (Fig.~\ref{fig:data}a). We look for a \emph{sparse texture}---a stretch of at least 3\,s in which no more than four stems play, in a combination that occupies less than a fifth of the track---and for an \emph{instrument departure}, where a stem that has been playing for at least 6\,s falls silent for at least 3\,s. When a sparse texture begins right after a departure, the same span can be described in two ways: a \emph{state} query saying what is playing (``only piano and bass are playing''), and a \emph{change} query saying what just left (``the guitar drops out, leaving piano and bass''). The \emph{change} query contains the full \emph{state} description plus the departure.

Pairing is what makes the two conditions comparable: both queries point to the same audio, position, and duration, so chance is the same by construction. This matters because the two kinds of moments are not distributed alike---sparse textures cluster heavily at track openings, where always predicting the start already reaches R@1@0.5 $=0.19$, whereas departures are spread across the track (0.01), while the evaluated models also show a preference for early positions. Pairing also reduces this positional bias, because spans near the track opening cannot follow a departure and are therefore excluded.

While the stem metadata provides precise event boundaries, we additionally filter for audibility in the rendered audio: an active stem must exceed $-6$ dB best-band SNR against the stem-subtracted mixture and remain within 20 dB of the track median. Fade-outs are removed and spans are capped at 30 s. After removing 221 duplicated tracks, this yields 713 pairs on 503 tracks. We hold out 20\% (143 pairs, stratified by position and template) as a dev split used only to fix two post-processing parameters and select the attention variant; the remaining 570 are evaluated. Queries use 13 coarse instrument classes, are paraphrased by an LLM with both members of a pair rewritten together, and pass checks for positional or duration words, length leakage, matched length distributions, and pair integrity (\emph{change} $\supseteq$ \emph{state} for all 713).

\vspace{-0.53em}
\subsection{Span extraction experiments}
\label{ssec:experiments}

Every experiment uses the same frozen model, audio, and query; only the way the span is extracted differs, each method asking a different question.

\para{Exp.~1 --- full-track prompting: \emph{can the model say when?}} The full track and the query are given with a fixed prompt asking for start and end seconds, and the generated text is parsed.

\para{Exp.~2 --- windowed yes/no scoring: \emph{does the model verify the moment when handed the segment, and does it need to hear what came before?}} We never ask for a time. We slide a window along the track and, at each position $t$, ask whether the final 2\,s of the window match the query; instead of letting the model answer, we read the next-token logits and take $s(t)=\mathrm{logit}[\text{yes}]-\mathrm{logit}[\text{no}]$. The model scores candidate segments rather than predicting their timestamps. The window $[t-C,\,t+2]$ always ends with the judged 2\,s segment and begins $C$ seconds earlier (Fig.~\ref{fig:data}b). The \textbf{preceding context} $C\in\{0,2,4,8\}$\,s is the one variable we manipulate. With $C{=}0$, the model hears only the judged segment, so the departure in a \emph{change} query cannot be verified---only its residual state content can. As $C$ gets longer, the departure enters the window. Whether \emph{change} queries recover with $C$ thus tests directly whether the model compares the segment with what preceded it. Scores are $z$-normalized per (track, query) across windows, removing each query's overall tendency to say yes. A question-free variant that instead scores the log-probability of the query text is examined in Sec.~\ref{ssec:r1r3}.

\para{Exp.~3 --- full-track attention: \emph{without any window scaffolding, does the model find the moment on its own?}} Exp.~3 asks whether the model can identify the relevant moment internally when given the full track, even if it cannot verbalize the timestamp. We read attention from the final prompt token to the audio tokens, averaging over the last third of layers and all heads. Layer, source position, and normalization were fixed in advance, and only models with a positive dev in-span attention excess (CI excluding 0) were evaluated.

\para{From score curves to spans.} Exp.~2 and Exp.~3 produce temporal score curves rather than span boundaries. For standard evaluation, we convert each curve to a span by expanding from its maximum while the score remains within a tolerance $\delta$, with a minimum duration; both parameters are fixed on the dev set. Because this heuristic can penalize short targets even when their location is correct, we also report an oracle-duration diagnostic that fixes the predicted duration to the ground-truth duration and evaluates localization alone.

\section{Experimental Setup}
\label{sec:setup}
\vspace{-0.83em}
\para{Models.} We evaluate seven open-weight LALMs: Qwen2-Audio-7B~\cite{chu2024qwen2audio}, Qwen2.5-Omni-3B~\cite{xu2025qwen25omni}, Qwen3-Omni-30B-A3B~\cite{qwen3omni2025}, MOSS-Audio-4B~\cite{yang2026mossaudio}, MOSS-Music-8B~\cite{mossmusic2026}, Audio Flamingo 3~\cite{ghosh2025af3}, and Music Flamingo~\cite{ghosh2026musicflamingo}. Selection required open weights, a known audio-token-to-time mapping, and full-track input in one pass (Qwen2-Audio's 8k context precludes Exp. 1 and 3), and within these constraints varied the encoder's \emph{segment length}. In four models, encoder self-attention is confined to a short window of the input: 2\,s blocks in Qwen2.5-Omni, and 1\,s (Qwen3-Omni) and 4\,s (MOSS) windows, which we set through the encoders' inference-window parameters. The other three use Whisper-style encoders that pad every input to 30\,s with absolute positional embeddings inherited from Whisper's 30\,s training windows~\cite{radford2023whisper}, so a long track is encoded as a sequence of independent 30\,s frames. The two MOSS models share an encoder and an input pipeline that interleaves time-marker tokens with the audio tokens; MOSS-Music adds music-specific continual pretraining and instruction tuning.

% Within these constraints we chose models so that the encoder's \emph{segment length} varies. Four models use encoders whose self-attention is bounded to a local window of the input—2-s blocks in Qwen2.5-Omni, an 8-s inference window over 1-s chunks in Qwen3-Omni's AuT, and an 8-s sliding window in the MOSS encoder—so each audio representation depends only on nearby audio; we call these short-segment encoders for brevity. The other three use Whisper-style encoders that pad every input to 30 s and apply absolute positional embeddings inherited from Whisper's 30-s training windows [22], so every representation is computed inside a fixed 30-s frame regardless of input length, and a long track is encoded as a sequence of independent 30-s frames. The two MOSS models share this encoder, and their input pipeline interleaves a time-marker token every 2 s of audio; MOSS-Music adds music-specific continual pretraining and instruction tuning.

% \vspace{-0.13em}
\para{Data.} All experiments use the same 570 evaluation pairs (1,140 queries); Exp.~2 uses a 1\,s hop for each $C$. Padded-encoder models score the truncated windows at a track's very start spuriously high (47--65\% of argmaxes in the first 8\,s at $C{=}8$, vs.\ 5--18\% otherwise), so windows whose judged segment begins at $t<8$\,s are excluded for all models and $C$, chosen after inspecting the argmax distribution of the padded models. No evaluation span begins there.
% \vspace{-0.13em}
\para{Metrics and chance.} We report R@1@0.3 for the standard spans, where the minimum-duration constraint can limit IoU, and R@1@0.5 for oracle-duration spans, where this boundary effect is removed. We use a best \emph{constant} span baseline rather than random span---the single normalized center and length that maximizes R@1 over the evaluation set, i.e., what a model achieves by always guessing the most common location (0.147 at IoU 0.3, 0.098 at 0.5, and 0.100 for oracle-duration).
% \vspace{-0.13em}
\para{Controls.} A high score could arise without the model actually matching this query to this audio: it might respond to the track alone, regardless of query; it might produce a peak for any query, because normalized scores are relative; or it might answer from position without listening. One control rules out each. \emph{Query shuffle} scores a track with the curve of a different query from the same track---if performance survives, the model reacts to the track, not the query. \emph{Cross-track} uses a query that describes nothing in this track---performance should fall to chance and the raw yes/no score should drop, showing that the score means something in absolute terms. \emph{Audio shuffle} permutes the track in 2\,s pieces before Exp.~1 and 3---if the answer does not change, it came from a positional prior, not the audio (Exp.~2 scores windows independently, so it does not apply).
% \vspace{-0.13em}
\para{Statistics.} To test whether differences across query types and context conditions are reliable, all comparisons are paired on the same span: McNemar's test on discordant pairs for R@1, with a bootstrap CI that resamples tracks rather than pairs, since pairs on one track are not independent. Comparisons across $C$ and \emph{state}–\emph{change} gaps are Holm-corrected, and whether a single cell exceeds chance uses the same track-resampled 95\% CI.

\section{Results}
\label{sec:results}

\begin{figure*}[t]
\centering
\includegraphics[width=0.75\textwidth]{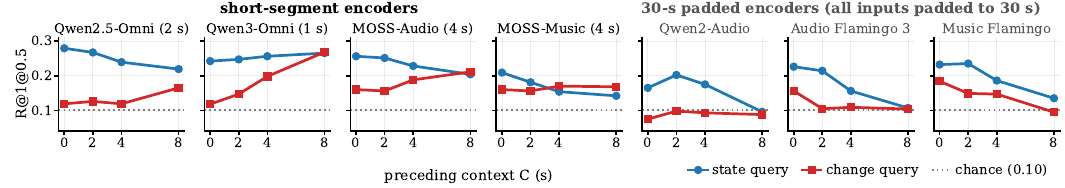}
\caption{Exp.~2, oracle-duration R@1@0.5 versus preceding context $C$ for state (blue) and change (red) queries on the 570 evaluation pairs. Change queries recover with context in three of the four short-segment models (left); the three 30 s padded models (right) degrade as $C$ grows regardless of training.}
\label{fig:groupC}
\end{figure*}

\begin{table*}[t]
\centering
\scriptsize
\setlength{\tabcolsep}{2.6pt}
\begin{tabular}{@{}l l cccc cccc cc c@{}}
\toprule
& & \multicolumn{4}{c}{Exp.~2, standard R@1@0.3} & \multicolumn{4}{c}{Exp.~2, oracle-duration R@1@0.5} & \multicolumn{2}{c}{Exp.~1 R@1@0.5} & Exp.~3 \\
\cmidrule(lr){3-6}\cmidrule(lr){7-10}\cmidrule(lr){11-12}\cmidrule(lr){13-13}
& & \multicolumn{2}{c}{State} & \multicolumn{2}{c}{Change} & \multicolumn{2}{c}{State} & \multicolumn{2}{c}{Change} & & & attn. excess \\
Model & Encoder segment & $C{=}0$ & $C{=}8$ & $C{=}0$ & $C{=}8$ & $C{=}0$ & $C{=}8$ & $C{=}0$ & $C{=}8$ & state & change & state / change \\
\midrule
Qwen2.5-Omni-3B & Whisper-v3 (block attn.) & 0.221$^{\ddag}$ & 0.240$^{\ddag}$ & 0.105 & 0.219$^{\ddag}$ & 0.279$^{\ddag}$ & 0.219$^{\ddag}$ & 0.119 & 0.165$^{\ddag}$ & 0.011 & 0.011 & $+$0.041 / $+$0.020$^{*}$ \\
Qwen3-Omni-30B-A3B & AuT & 0.247$^{\ddag}$ & 0.298$^{\ddag}$ & 0.102 & \textbf{0.312}$^{\ddag}$ & 0.242$^{\ddag}$ & 0.265$^{\ddag}$ & 0.118 & \textbf{0.268}$^{\ddag}$ & 0.042 & 0.025 & $+$0.044 / $+$0.062$^{*}$ \\
MOSS-Audio-4B & MOSS Audio Encoder $^{\dagger}$ & 0.232$^{\ddag}$ & 0.237$^{\ddag}$ & 0.163 & 0.233$^{\ddag}$ & 0.256$^{\ddag}$ & 0.204$^{\ddag}$ & 0.160$^{\ddag}$ & 0.211$^{\ddag}$ & 0.030 & 0.019 & n.s. \\
MOSS-Music-8B & MOSS Audio Encoder $^{\dagger}$ & 0.223$^{\ddag}$ & 0.184$^{\ddag}$ & 0.139 & 0.170 & 0.209$^{\ddag}$ & 0.142$^{\ddag}$ & 0.160$^{\ddag}$ & 0.168$^{\ddag}$ & \textbf{0.072} & 0.042 & n.s. \\
Qwen2-Audio-7B & Whisper-v3 (30 s pad.) & 0.181$^{\ddag}$ & 0.125 & 0.096 & 0.111 & 0.165$^{\ddag}$ & 0.096 & 0.075 & 0.088 & --- & --- & --- \\
Audio Flamingo 3 (7B) & AF-Whisper (30 s pad.) & 0.230$^{\ddag}$ & 0.142 & 0.147 & 0.107 & 0.226$^{\ddag}$ & 0.107 & 0.156$^{\ddag}$ & 0.104 & 0.007 & 0.004 & n.s. \\
Music Flamingo (7B) & AF-Whisper (30 s pad.) & 0.214$^{\ddag}$ & 0.135 & 0.140 & 0.114 & 0.232$^{\ddag}$ & 0.135$^{\ddag}$ & 0.184$^{\ddag}$ & 0.095 & 0.012 & 0.009 & n.s. \\
\midrule
Chance (best constant span) & & 0.147 & 0.147 & 0.147 & 0.147 & 0.100 & 0.100 & 0.100 & 0.100 & 0.098 & 0.098 & 0 \\
\bottomrule
\end{tabular}
\caption{All results on the 570 evaluation pairs. Exp.~2: R@1@0.3 with standard post-processing and R@1@0.5 with oracle duration at $C{=}0$/$8$; Exp.~1: R@1@0.5; Exp.~3: span-specific in-span attention excess ($z$-units). $^{\ddag}$95\% track-bootstrap CI lower bound exceeds chance (2,000 resamples, uncorrected; no Exp.~1 cell qualifies). $^{*}$CI excludes 0 for both conditions; n.s.: CI includes 0; $^{\dagger}$Input pipeline interleaves 2\,s time markers.}
\label{tab:main}
\end{table*}
\vspace{-0.33em}

\subsection{State queries can be verified}
\label{ssec:state}

Handed the segment alone ($C{=}0$), every model places \emph{state} queries above the baseline through windowed scoring under both standard and oracle-duration evaluation (Table~\ref{tab:main}, Fig.~\ref{fig:groupC}). In every model the \emph{state} query also beats the \emph{change} query on the very same spans (Holm $p<0.03$). Four models also place \emph{change} queries above chance under oracle duration at $C{=}0$, consistent with their embedded state description, but none reaches the \emph{state} level; under standard post-processing, none has a CI over chance. The controls confirm a genuine query--audio match: a shuffled or cross-track query drops performance to near zero, and a cross-track query's raw yes/no score is significantly lower than the true query's in all but three of the results.

\vspace{-0.63em}
\subsection{Change queries need preceding context}
\label{ssec:context}

Whether a \emph{change} query becomes answerable once the model hears what came before differs by model, and the difference follows the encoder. In three of the four short-segment models, \emph{change} queries rise with $C$ (Fig.~\ref{fig:groupC}, left): Qwen2.5-Omni, MOSS-Audio, and above all Qwen3-Omni, whose \emph{change} queries go from near chance at $C{=}0$ to the level of its \emph{state} queries at $C{=}8$ (within-condition $p=2\!\times\!10^{-19}$; gap n.s., Holm $p=0.92$). The gap also closes for MOSS-Audio ($p=0.80$) and shrinks to a third for Qwen2.5-Omni ($p=0.003$). Although \emph{state} scores decline mildly with $C$ in two of the three, \emph{change} improves significantly in all three, and these are the only models whose $C=8$ \emph{change} scores clear chance at IoU 0.3. Qwen3-Omni shows the strongest recovery; its larger scale is consistent with this pattern, although scale is confounded with encoder design. MOSS-Music is the exception among short-segment models with its \emph{change} score rising only slightly and non-significantly ($p=0.09$), while its \emph{state} score declines, accounting for most of the gap reduction. Because MOSS-Music shares its encoder with MOSS-Audio, this divergence shows that the encoder does not fully determine context use: training can suppress a recovery that the encoder permits.

The three remaining models share Whisper-style encoders that pad every input to 30\,s and, despite different language models and training, behave alike (Fig.~\ref{fig:groupC}, right): \emph{change} queries show no meaningful recovery, \emph{state} queries degrade toward chance as $C$ grows (Audio Flamingo 3 loses half its state score), and at IoU 0.3 all three fall below chance for both conditions at $C{=}8$. (State--change gap at $C{=}8$: Holm $p=0.54$, $0.90$, $0.013$.) One possible explanation is that padding hides where the audio ends: at $C{=}0$ the audio is just the judged 2\,s, but as $C$ grows the model must isolate those 2\,s inside a 30\,s frame, so its judgment drifts toward the whole window. This shared pattern across differently trained models suggests an encoder-related effect, although the evidence remains correlational and may not be causal.

\subsection{Prompting and attention struggle to expose the signal}
\label{ssec:r1r3}

\para{Prompting.} Asked for timestamps directly, all models evaluated in Exp.~1 are at or below chance, and on the same pairs windowed scoring is far better. The prompted answers are largely independent of the audio: most models return a constant that depends only on the query's shape (a track-opening span for \emph{state}, a mid-track one for \emph{change}), and for four models most answers survive audio shuffling unchanged. The exception is MOSS-Music: its answers do change under shuffling and their centers land inside the true span four times more often than random, but its spans are about one second long, so IoU still fails. 

\para{Scoring variants.} Two variants of Exp.~2 clarify what the logit scoring preserves. Verbalizing each window-level judgment as yes/no sharply degrades performance—halving \emph{state} and pushing \emph{change} below chance—by discarding the ranking carried by the yes–no logit margin. Removing the judgment prompt and scoring only the query also stays at chance across models, so the useful signal lies in the graded margin of an explicit but ungenerated judgment.

% \vspace{-0.95em}
\para{Attention.} We measure the \emph{in-span attention excess}---mean normalized attention inside the true span minus outside, in $z$-units---and its \emph{span-specific} version, the excess for the target query minus that for another query on the same track. Only the two Qwen-Omni models show significant span-specific excess, but even there span localization remains below chance. Under audio shuffling the attention follows the content to its new location rather than staying at the original position, indicating local acoustic matching rather than reliable track-level localization. This is consistent with Exp.~2: verifying a segment that is explicitly presented does not imply that the model can locate the same segment within the full track. The model may hear the segment when handed it, yet fail to find where it was heard; Exp.~2 supplies the search, whereas Exp.~3 asks the model to perform it.

\vspace{-1.63em}
\section{Discussion and Limitations}
\label{sec:disc}
\vspace{-0.33em}
Windowed scoring reveals a consistent distinction between being able to verify a candidate segment and being able to locate it in a full track. Across models, local states are recoverable when the relevant audio is explicitly presented, while verifying the change itself benefits from preceding context. Full-track prompting and attention expose much less of this signal, suggesting that temporal grounding failures in LALMs are not explained solely by a simple absence of temporal information, but also by how that information is accessed and searched over time.

The context results further suggest that this access is shaped, though not fully determined, by the audio encoder. Short-segment encoders are generally able to benefit from additional context, whereas others show a shared degradation pattern despite differences in language model and training. Music-specific fine-tuning does not consistently alter this behavior, which suggests that encoder-side temporal organization may matter alongside model scale and domain specialization.

These findings should be interpreted within the scope of the probe. Slakh2100 provides precise stem activity, but it is MIDI-rendered and our queries are restricted to instrumentation, so \emph{state} and \emph{change} capture only which instruments are present or disappear, not other temporally meaningful musical properties such as harmony, dynamics, rhythm, or vocal structure. Extending the probe to real recordings with vocals, for example using MUSDB18, would therefore test whether the same patterns hold under less controlled but more realistic acoustic conditions.

The paired design makes the comparisons well controlled, but the absolute scores remain modest across methods. This leaves substantial room for work that targets music temporal grounding more directly, including richer query types, more realistic recordings, and methods that bridge the gap between local verification and full-track localization.

\vspace{-1.53em}
\section{Conclusion}
\label{sec:conc}
\vspace{-0.50em}
We asked whether frozen large audio-language models fail at music temporal grounding because temporal information is absent, or because standard prompting does not expose it. Using a training-free paired probe built from Slakh2100 and evaluated on seven open-weight LALMs, we find that models can verify local musical states and, in several short-segment models, context-dependent changes even when full-track prompting and attention fail to localize them reliably. These results suggest a distinction between verifying that a candidate segment matches a query and locating that segment within a full track. Prompted timestamp failure therefore does not fully reflect the temporal information accessible to these models, while reliable full-track localization remains limited.

\vfill\pagebreak
\bibliographystyle{IEEEbib}
\bibliography{refs}

\end{document}